\documentclass[a4paper,fleqn,usenatbib]{mnras}

\usepackage[T1]{fontenc}
\usepackage{ae,aecompl}

\usepackage{graphicx}	
\usepackage{amsmath}	
\usepackage{amssymb}	
\usepackage{xcolor}

\title[A Pre-Merger Stage Galaxy Cluster: Abell 3733]{A Pre-Merger Stage Galaxy Cluster: Abell 3733}

\author[H. \.Ilker KAYA et al.]{
H. \.Ilker KAYA,$^{1}$
Turgay CAGLAR,$^{1,2}$\thanks{E-mail: caglar@strw.leidenuniv.nl}
Hakan SERT$^{1,3}$
\\
$^{1}$Y{\i}ld{\i}z Technical University, Faculty of Science and Art, Department of Physics, Istanbul 34220, Turkey\\
$^{2}$Leiden University, Leiden Observatory, Astronomy Department, Leiden, 2380RA, The Netherlands\\
$^{3}$von Karman Institute for Fluid Dynamics, Chauss\'ee de Waterloo, 72, B-1640 Rhode-St-Gen\'ese, Belgium
}
\date{Accepted XXX. Received YYY; in original form ZZZ}

\pubyear{2019}
\volume{485}

\begin{document}
\label{firstpage}
\pagerange{4550-4558}
\maketitle

\begin{abstract}

The galaxy cluster Abell 3733 (A3733) is a very suitable candidate in addressing dynamical processes throughout galaxy cluster mergers. This study shows structural analysis results of A3733 (z = 0.038) based on X-ray and optical data. According to X-ray luminosity map, A3733 hosts two sub-structures separated in the sky by $\sim$ 0.25 Mpc, and the two distinct clumps are located in the East (A3733E) and the West (A3733W) directions. Both sub-structures are centred on two different brightest cluster galaxies (BCGs), and the X-ray and optical centroids of both BCGs substantially coincide with each other. The intracluster medium (ICM) temperatures of the sub-structures are estimated to be 2.79 keV for A3733E and 3.28 keV for A3733W. Both sub-structures are found to be hosting cool central gas (kT $ \sim $ 1.5-2.5 keV) surrounded by hotter gas (kT $\sim $ 3.0-3.5 keV). Besides, the X-ray concentration parameters are found to be c $\sim $ 0.3 for each sub-structure. These results indicate the existence of cool centres for both sub-structures. The optical density map reveals a crowded galaxy population within the vicinity of A3733W. The high probable (\% 88.2) dynamical binding model of A3733 suggests that the cores of sub-structures have a 3D separation of 0.27 Mpc and will collide in 0.14 Gyr with the relative in-falling velocity of 1936 km s$^{-1}$. As a conclusion, this study demonstrates some evidence suggesting that the A3733 system is in the pre-merger state. 

\end{abstract}

\begin{keywords}
X-rays: galaxies: clusters -- galaxies: clusters: general -- galaxies: clusters: individual: Abell 3733 -- galaxies: clusters: intracluster medium
\end{keywords}




\section{Introduction}

Clusters  of  galaxies  are  the  largest cosmic laboratories in which evolution of the universe can be studied in detail. They are formed by virtue of in-falling of the other objects through a central dominant object by the gravitational attraction \citep{Press}. One of the most crucial results of gravitational attraction is galaxy clusters mergers, which are the most energetic ($\sim$ 10$^{64}$ ergs) events in the universe \citep{Sarazin}. Numerical simulations also  demonstrate that galaxy clusters are formed via accretion of sub-clusters along the filaments \citep[e.g.,][]{West}. Therefore, more detailed investigations of merging systems are essential for understanding the structural formation of the universe. 

The  theoretical  concept  of mergers is the angular momentum and  energy  transfer  between  the  merging  galaxy  clusters \citep[e.g.,][]{Sarazin}. The ICM  temperature of vicinity of merging systems can be increased as a result of galaxy cluster interactions. Mergers  are able to stop the cooling flow and can also destroy (or relocate) cool centres of galaxy clusters \citep{Markevitch2000,Markevitch2007}. 

Optical and X-ray correlations have been very successful for explaining the dynamics of mergers. The optical and Xray sub-clustering are found to be well-correlated for the majority of binary systems \citep[e.g.,][]{Baier,Kolokotronis}. X-ray investigations of the binary systems reported the existence of hot regions in-between sub-structures for many samples \citep{Gutierrez,Sarazin2013,Kato2015,Akamatsu2016,Akamatsu2017,Caglar1,Caglar2,Botteon18,Hallman}. But, the origin of hot region between the merging sub-structures was proposed by the different mechanisms, such as shock-heating or adiabatic compression \citep[][]{T99,Zuhone,Sch}. On the other hand, some merging binary galaxy clusters with higher separations did not show a strong evidence for shock heating \citep[e.g.,][]{Fujita1996,Fujita2008,Werner2008}. The extended radio emissions, such as radio halo and relics, are proposed to be generated by the turbulence after the collisions between the cores; therefore, they are not observed in the pre-merging binary galaxy clusters \citep{book,Feretti}.

X-ray analysis of A3733 was performed with ROSAT data by \citet{Ebeling1996} resulting in kT = 2.2 keV and L$_{x}$ = 4.2$\times$10$^{43}$ erg s$^{-1}$. Then, \citet{Piffaretti} used ROSAT data to measure cluster's mass and luminosity at the overdensity of 500 and provided the following results: r$_{500}$ = 678.9 kpc, L$_{500}$ = 2.8$\times$10$^{43}$ erg s$^{-1}$, and M$_{500}$ = 9.2$\times$10$^{13}$ M$_{\odot}$. A3733 is a Bautz-Morgan type I-II galaxy cluster with richness class R = 1 (\citet{RoRo}) and redshift z = 0.038 \citep{Dalton1994,SoSt,Smith2004}. The kinematic investigations of A3733 were performed by \citet{Stein1996,Stein1997} and \citet{SoSt}; however, neither result demonstrated significant sub-structure within the cluster's potential well. NGC 6999 is classified as the brightest cluster galaxy (BCG) of A3733 by \citet{PoLa}, and in the same work, the second brightest galaxy of A3733 is reported to be NGC 6998 with a slight absolute magnitude difference $\Delta$M = 0.013.  

This study, which reports an optical and X-ray investigation of merging galaxy cluster A3733, aimed to understand the physical structure of A3733 using X-ray and optical comparison. Dynamical events of A3733 occurred by mergers are also discussed in the present study, which is organised in the following manner: Section 2 presents the observation logs and data processing; Section 3 describes X-ray and optical data analysis procedures; Section 4 explains the Newtonian gravitational binding criterion of two-body systems;  Section 5 covers the discussion of our results and finally, we summarise our results in section 6. We adopt in this paper a standard $\Lambda$CDM cosmology parameters: H$_{0}$ = 70 km s$^{-1}$ Mpc$^{-1}$, $\Omega$$_{M}$ = 0.3 and $\Omega$$_{\Lambda}$ = 0.7 for a flat universe. In this cosmology, 1$\arcmin$ corresponds to 44.48 kpc. Unless stated otherwise, the error values are quoted at the \% 90 confidence interval in our analysis. 


\section{Observations and Data Processing}
The first \textit{XMM-Newton} observation was performed on 2015 November 30 for exposure of 15.8 ks, and the second \textit{XMM-Newton} observation was performed on 2016 December 8 for exposure of 18.5 ks. The medium filter was used for MOSs and pn cameras on the both observations. X-ray observations were taken in full frame for MOSs and extended full frame for pn. Suzaku satellite was also used to observe A3733 with 3 x 3 clocking mode on 2016 August 16 for exposure of 14.6 ks. The observation was performed with two front illuminated (FI) CCD chips (XIS 0 and XIS 3) and one back illuminated (BI) CCD chips (XIS 1). X-ray observational data were gathered from \textit{XMM-Newton} Science Archive (XSA) and Suzaku Data Archive and Transmission System (DARTS). The log of observations are presented in Table \ref{obslog}. The first radio observation was performed at a frequency of 4850 Mhz on the Parkes-MIT-NRAO (PMN) survey \citep{Wright}. 1400 Mhz radio observation of A3733 was performed on the The $NRAO$ $VLA$ Sky Survey (NVSS) \citep{Condon}, and the last observation at the frequency of 150 Mhz was performed on the the $GMRT$ all-sky radio survey \citep{Intema}. Radio images were gathered from Skyview archive. 

For \textit{XMM-Newton} data, we performed data reduction using \textit{XMM-Newton} Science Analysis Software (\textit{XMM-SAS v15.0} and \textit{XMM-Newton} Extended Source Analysis Software (\textit{XMM-ESAS}). Current calibration file (ccf) and summarised observation data file (odf) were created using the tasks: cifbuild-4.8 and odfingest-3.30, respectively. The emchain-11.19 and epchain-8.75.0 tasks were applied to data, which generated MOS and pn event files, respectively. The light curve was generated to determine a final net data, and corrupted data  was extracted performing evselect-3.62. Finally, point-like X-ray sources within the galaxy cluster were detected using edetect\_chain-3.14.1, and the point sources are removed from data in our analysis.
Suzaku data analysis were processed with heasoft version 6.21 and the latest calibration database (CALDB-2014-05).

\begin{table}
\begin{center}
\caption{X-ray Observations.}
\begin{tabular}{@{}cccl@{}} 
\hline
\hline
ObsID	    	& 	Satellite		& 	Date Obs		& Exposure   	\\
            		&                   		&               		&  (ks)          	 		\\
\hline
809108010 	& \textit{Suzaku}	& 2016-08-16		& 14.6			\\
0765000301      & \textit{XMM-Newton} & 	2016-12-08    	& 18.5 	  			\\
0741580801  	&\textit{XMM-Newton} & 2015-11-30    	& 15.8 				\\
\hline
\end{tabular}
\label{obslog}
\end{center}
\end{table}


\section{Analysis}

\subsection{Optical Analysis}

\begin{figure}
 \begin{center}
 \includegraphics*[width=8.6cm]{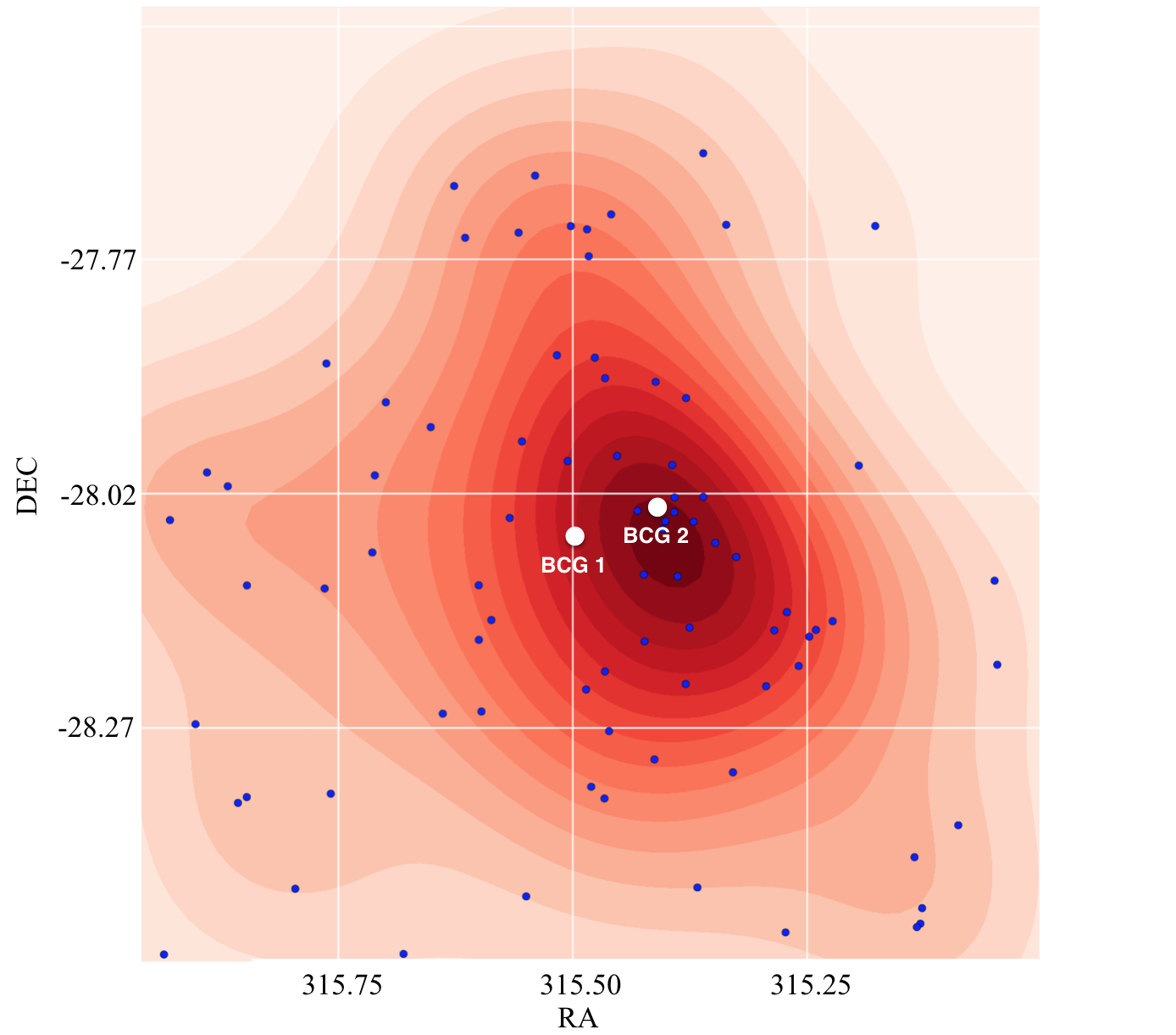}
  \includegraphics*[width=8cm]{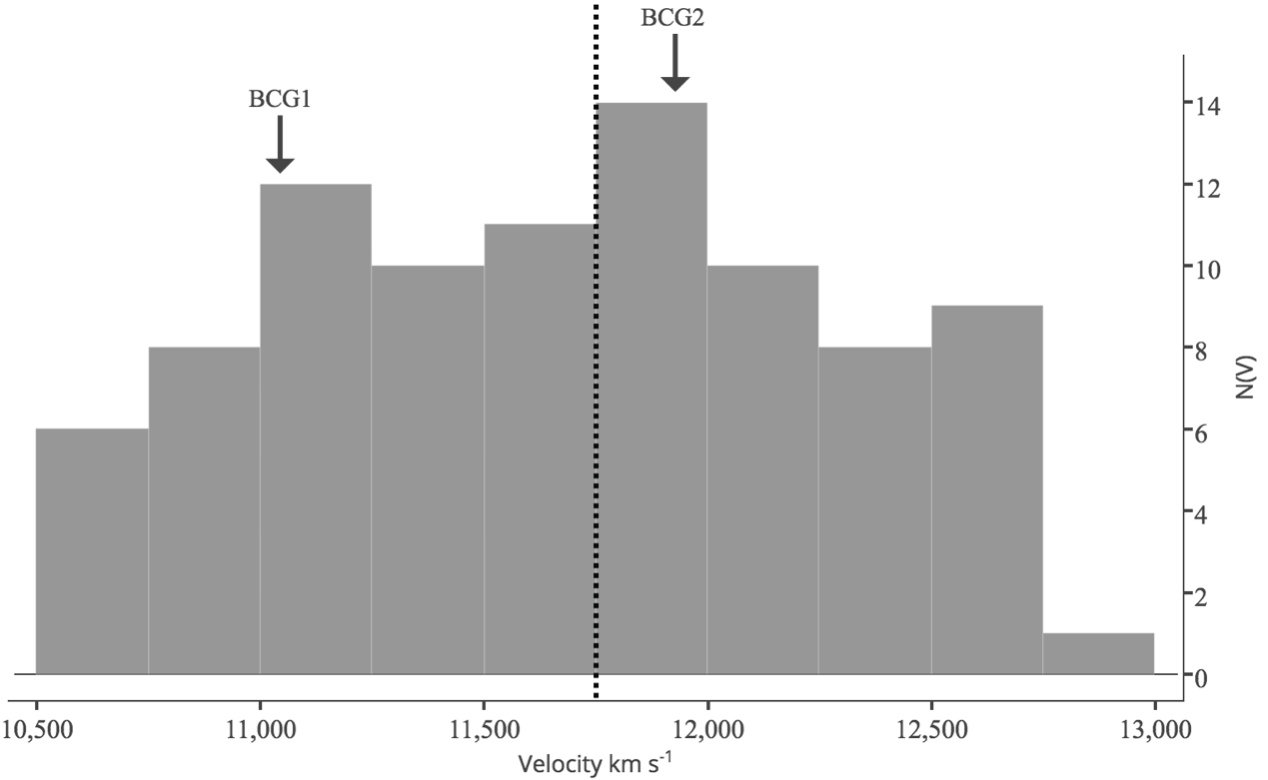}
 \caption{\label{optical-density} \textbf{Top:} The projected galaxy density map of A3733 within 1.5 Mpc radius of the field. The white and blue dots represent the position of BCGs and member galaxies, respectively. \textbf{Bottom:} Velocity histogram of A3733 (binning of 250 km s$^{-1}$) for 89 member galaxies. The black dashed line represents the mean velocity of A3733.}
\end{center} \end{figure}

\begin{figure}
\begin{center}
\includegraphics*[width=8.6cm]{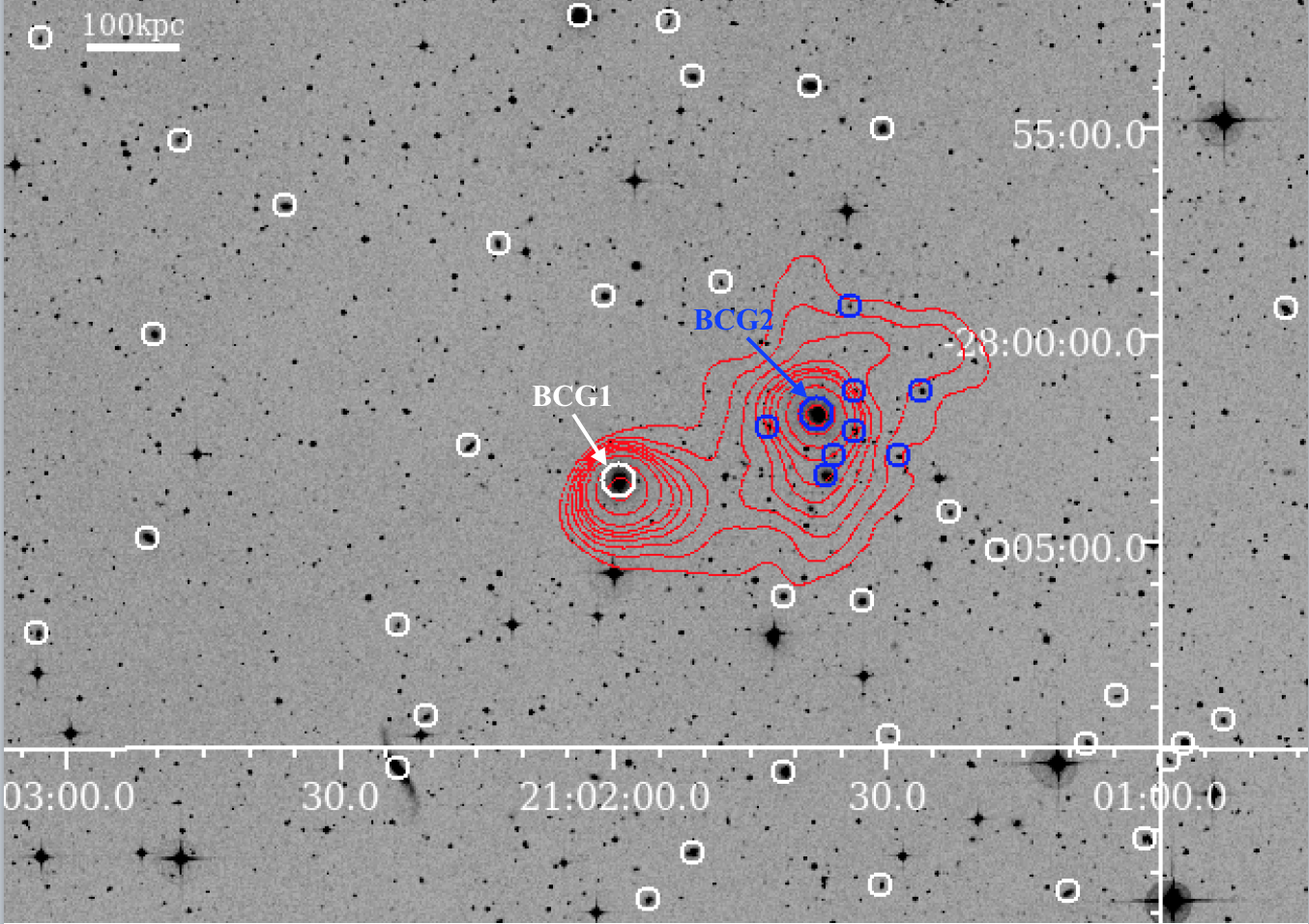}
\caption{\label{optic} Optical DSS image of A3733, overlaid with X-ray contours. X-ray contours are generated using background 
subtracted, exposure and vignetting corrected, combined $\textit{XMM-Newton}$ image. The white and blue circles represent the member galaxies of A3733. Member galaxies within the ICM of A3733W are presented as blue circles for visual aid.}
\end{center}
\end{figure}

We obtained spectroscopic redshifts of 89 galaxies within 1.5 Mpc radius of A3733 in the literature \citep{Stein1996,Katgert,SoSt,Jones}. Cluster membership were assigned with heliocentric velocities in the interval 10500 < V < 13000 km s$^{-1}$. Cluster member galaxies were used to generate the projected galaxy density map of A3733, which is presented in Fig. \ref{optical-density} (\textbf{top)}. The white and blue dots represent the position of BCGs and member galaxies, respectively. Moreover, we investigated the velocity histogram of A3733 using the cluster member galaxies (see Fig. \ref{optical-density} (\textbf{bottom}). Cluster's mean redshift is demonstrated as a black dashed line, and the velocities of BCGs are pointed with the black arrows for visual aid. We also present the member galaxies of A3733 in Fig. \ref{optic}, which clearly shows the a crowded region around BCG2. 

\subsection{Spatial Analysis}

\begin{figure*}
 \begin{center}
 \includegraphics*[width=8.4cm]{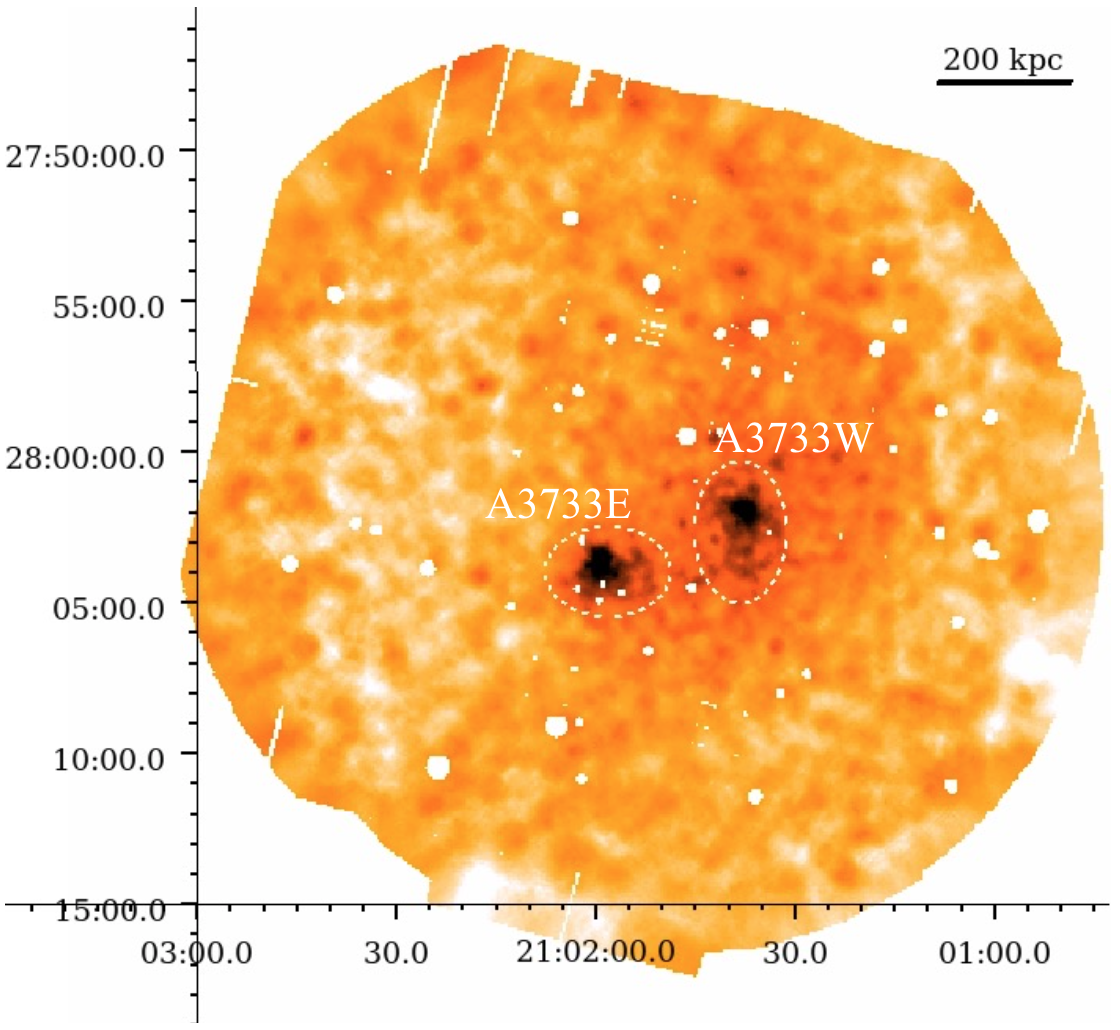}
  \includegraphics*[width=8.4cm]{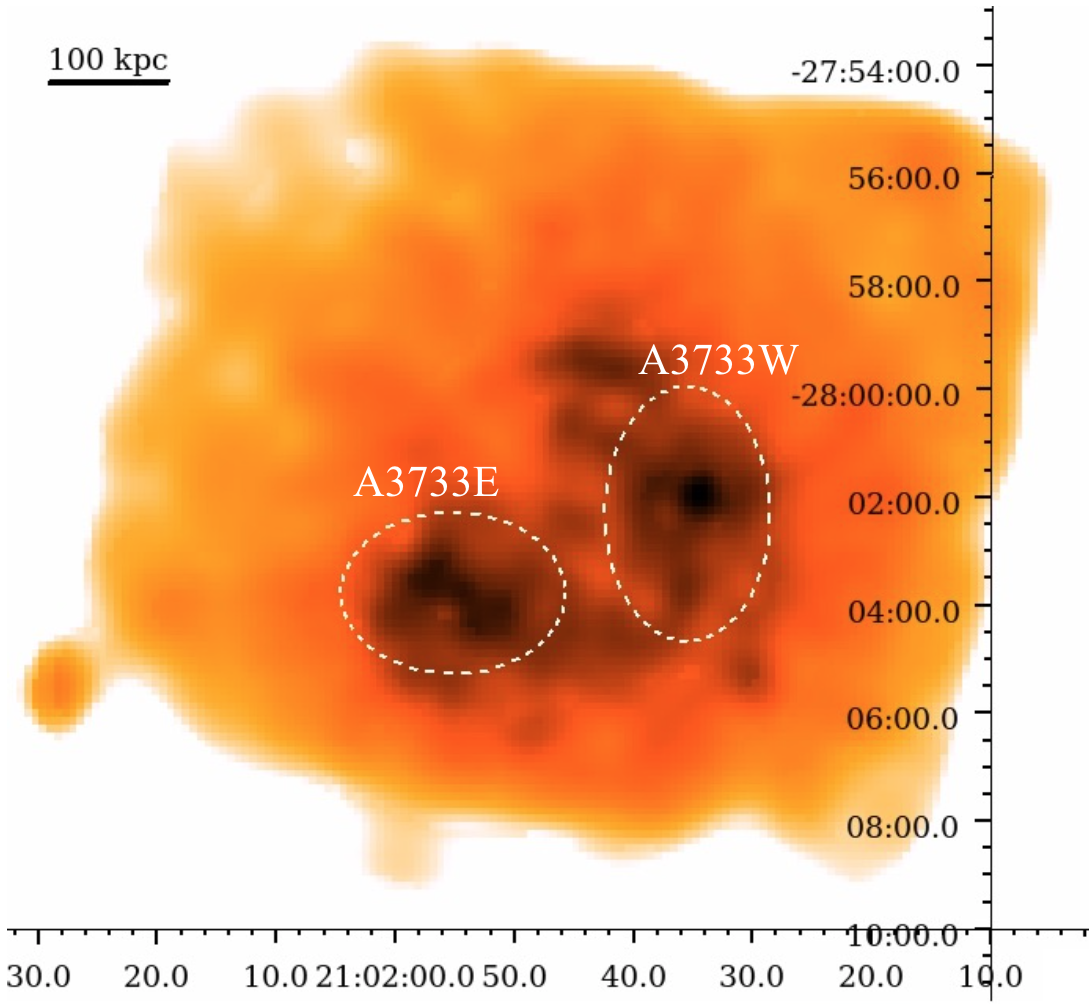}
 \caption{\label{back-sub-im} \textbf{Left:} The background subtracted, exposure and vignetting corrected, combined, and adaptive smoothed X-ray image within the energy range of 0.3 - 10 keV. \textbf{Right:} The combined and adaptive smoothed Suzaku raw X-ray image.}
\end{center} \end{figure*}

    The \textit{XMM-Newton} background subtracted, exposure and vignetting corrected, combined X-ray image was generated by the following analysis  procedure. MOS and pn raw data were created by the mos-filter and pn-filter task respectively. For MOS and pn using mos-spectra and pn-spectra, \textit{XMM-Newton} X-ray images were generated in the 0.4-10.0 keV energy range respectively. The point-like X-ray sources in A3733 were determined using the cheese task on the 5 $\times$ 10$^{-15}$ erg cm$^{-2}$ s$^{-1}$ flux threshold. The cheese task provides the event, exposure and mask images that are used for excluding point-like sources for the further purposes. In our analysis, point-like sources are excluded from the data by using mask parameter, which was set to 1.Proton scale, proton tasks were used  for removing soft proton and background contamination from the data. Since the \textit{XMM-ESAS} performs analysis within the detector coordinates, rot-im-det-sky task was used to convert the detector coordinates to celestial coordinates.
Finally, MOS and pn data were combined by comb task, and the combined image was smoothed adaptively by the adapt task. The background subtracted, exposure and vignetting corrected, combined, and adaptive smoothed X-ray image is presented in Fig. \ref{back-sub-im}.      

The surface brightness is defined as projected plasma emissivity per area on the sky. The $\beta$ model was used to fit the X-ray surface brightness of each sub-structures. \citep{Cavaliere}. The model is defined as: 

\begin{equation}
S (r) = S_{0} \times \left[ 1 + \left(r/r_{c}\right)^{2} \right]^{-3\beta+0.5} + c ,
\end{equation}
where S$_{0}$ is the central surface brightness, r$_{c}$ is the core radius, and $\beta$ is the shape parameter. In this equation, c parameter was added to estimate the background level, and it is crucial to estimate best-fit parameters of the surface brightness. Whereas the background level is found to be 2.22$\pm{0.22}$ $\times$ 10$^{-2}$ counts arcsec$^{-2}$, the resulting $\beta$ model parameters are presented in Table \ref{surbri}.

\begin{table}
\caption{\label{beta} The best-fit parameters of \textit{XMM-Newton} data for $\beta$-model and $r_{2500}$.}
\resizebox{0.58\textwidth}{!}{\begin{minipage}{0.49\textwidth}
\begin{tabular}{cccc} 
\hline
\hline
Region			&     $r_{c}$	& 			$\beta$  			&	 $r_{2500}$ 		\\	
				& 	(kpc) 			&				& 	 		(kpc) 		\\
\hline
A3733E	 	&	8.27$^{+2.79}_{-2.55}$	&	0.46$^{+0.065}_{-0.079}$	& 	429.8$^{+54.6}_{-47.7}$	\\
A3733W 		& 	6.98$^{+3.19}_{-2.50}$   	&	0.47$^{+0.067}_{-0.081}$	&	467.3$^{+63.4}_{-53.7}$	\\
\hline
\end{tabular}
\label{surbri}
\end{minipage}}
\end{table}

\subsection{Spectral Analysis}
The evselect-3.62  task was used to generate  spectrum  and  background  files for the spectral analysis of \textit{XMM-Newton}. The response files for the A3733 galaxy were generated using the rmfgen-2.2.1 and arfgen-1.92 tasks. The local background was removed from the source file using an annular region within 11$\arcmin$ - 12$\arcmin$ away from the cluster center. The instrumental background lines such as Al K-$\alpha$, Si K-$\alpha$ and Cu K-$\alpha$ were carefully removed from the data. The spectrum files of \textit{Suzaku} was created by XSELECT-2.4. Non X-ray background (NXB) was subtracted from spectra by the background file, which was generated by xisnxbgen-2010-08-22. The Cosmic X-ray background (CXB) were modelled with a photon index of 1.4 \citep{hickox}. The xissimarfgen-2010-11-05 and xisrmfgen-2012-04-21 tasks were used to generate the response files. All generated spectral files were grouped by grppha. We perform the spectral analysis of A3733 using XSPEC-12.9.1 \citep{Arnaud96}. Thermal model APEC \citep{Smith} and the X-ray absorption model \citep{Wilms} TBABS were used in our analysis to estimate plasma temperature. The \textit{XMM-Newton} data were simultaneously fitted within the energy range of 0.3 - 10.0 keV using 4 MOS and 2 pn data from both \textit{XMM-Newton} observations. The \textit{Suzaku} data was also fitted simultaneously within the energy range of 0.8 - 7.0 keV  using XIS0, XIS1 and XIS3 data.  The spectral fit results of A3733 are presented in Table \ref{spectable}.

\begin{table}
\begin{center}
\caption{\label{spectable}The spectral best-fit parameters of A3733 with $^1$\textit{XMM-Newton}  and $^2$\textit{Suzaku} data.}
\begin{tabular}{@{}l c c r@{}} 
\hline
\hline
Region			& 	$kT$					& Abundance  				& $\chi^2/dof$ 			\\	
				& 	(keV) 				& ($Z_{\odot}$) 			& 	 		 		\\
\hline
A3733E$^{1}$	 	&	2.79$^{+0.21}_{-0.22}$	&	0.38$^{+0.14}_{-0.12}$	& 	2066/1831 = 1.12		\\
A3733W$^{1}$ 		& 	3.28$^{+0.27}_{-0.25}$   	&	0.34$^{+0.13}_{-0.11}$	&	2058/1856 = 1.11	\\
Bridge$^{1}$ 		& 	4.14$^{+0.40}_{-0.51}$	& 	0.34$^{+0.14}_{-0.13}$	&	1544/1350 = 1.14		\\ 
A3733E$^{2}$ 		& 	2.49$\pm{0.41}$ 		&  	0.36$^{+0.21}_{-0.15}$	&	297/351 = 0.85		\\
A3733W$^{2}$ 		& 	3.12$^{+0.46}_{-0.41}$  	&	0.27$^{+0.23}_{-0.19}$	&	306/343 = 0.89		\\
Bridge$^{2}$ 		& 	4.09$^{+1.29}_{-0.95}$	& 	0.3 (fix)				&	136/152 = 0.89		\\ 
\hline
\end{tabular}
\end{center}
\end{table}

\begin{figure}
 \begin{center}
 \includegraphics*[width=7.6cm]{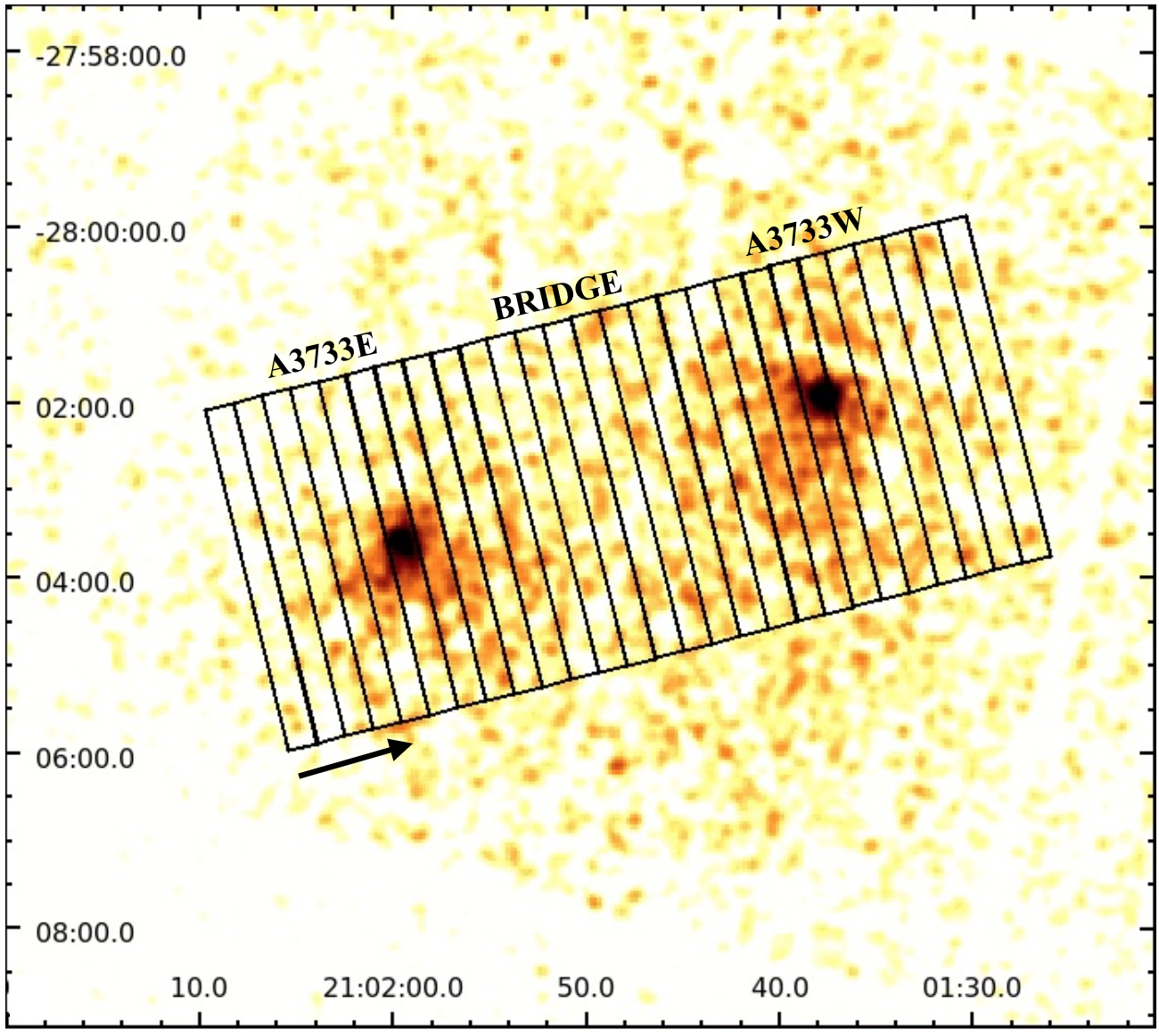}
 \includegraphics*[width=7.7cm]{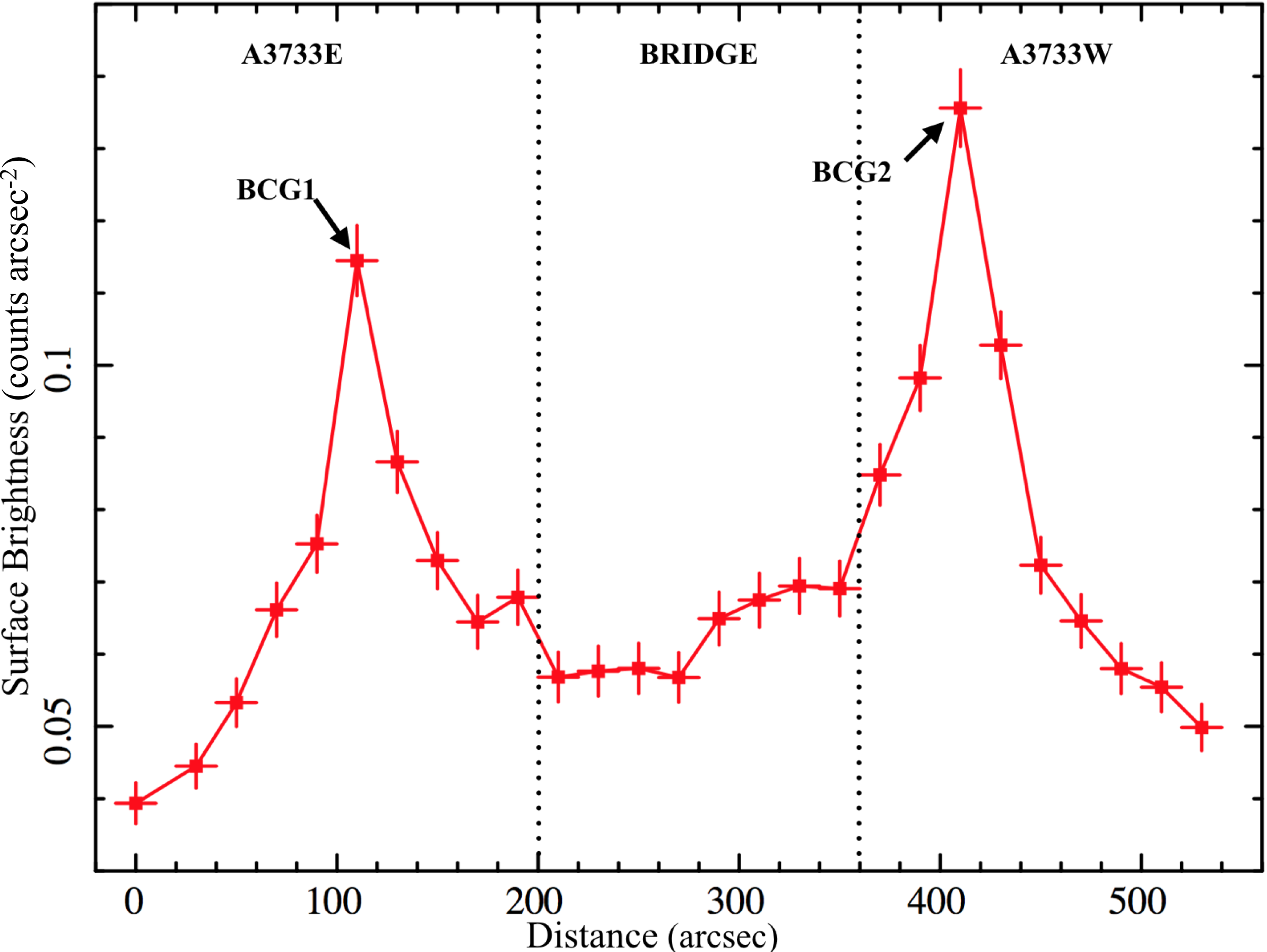}
 \caption{\label{sur-bri-im} The surface brightness profile of the A3733 system demonstrating the absence of compressed gas between sub-structures. }
\end{center} \end{figure}


\subsection{X-ray Morphological Parameters}

X-ray morphological parameters are essential tools to resolve dynamical disturbance level of galaxy clusters. To understand the disturbance of each sub-structure of A3733, we estimated the centroid shift parameters for each one. In undisturbed galaxy clusters, the centroid shift are expected to be spherically symmetric and roughly equal to zero. We used the method provided by \citet{Mohr} in our centroid shift estimations. The centroid shift parameter can be obtained by the following equation:

\begin{equation}
w = {\Bigg[ \frac{1}{N-1} \sum (\Delta_i - \langle \Delta \rangle)^2 \Bigg]}^{1/2} \times \frac{1}{R_{ap}},
\end{equation}
where $N$ is the total aperture number, $\Delta$ is the separation between X-ray peak and the centroid of $i$th aperture, and R$_{ap}$ is 500 kpc, which is decreased in steps of 5\%. 

The concentration degree of X-ray gas can be calculated from X-ray concentration parameter, which is highly essential in discriminating cool and non-cool core clusters. We estimated X-ray concentration parameters for each sub-structures by using the following equation provided by \citet{Santos}. 

\begin{equation}
c = \frac{S(r < 100 kpc)}{S(r < 500 kpc)},
\end{equation}
where $S$ represents the surface brightness within 100 kpc and 500 kpc, respectively. We present the resulting X-ray morphological parameters in Table \ref{morpho}

\begin{table}
\caption{\label{morpho} X-ray morphological parameters of A3733 obtained from $XMM-Newton$ data. MD = Moderately Disturbed, CC = Non-cool core.}
  \resizebox{0.58\textwidth}{!}{\begin{minipage}{0.46\textwidth}
\begin{tabular}{cccc} 
\hline
\hline

Cluster			& 		w					&			 c  	& Note \\	
\hline
A3733E 		& 		0.059$\pm{0.002}$			&  0.33$\pm{0.02}$ & MD-CC \\
A3733W 		& 		0.037$\pm{0.001}$			&  0.29$\pm{0.02}$ & MD-CC \\

\hline
\end{tabular}
\end{minipage}}
\end{table}

\subsection{Galaxy Cluster Mass Calculations}
In hydrostatic equilibrium and isothermal spherical symmetry, galaxy cluster total mass can be estimated by adopting $\beta$ model parameters \citep{LimaNeto}:
\begin{equation}
    M(r) =  \frac{3 k T_{0} \beta r_{c}}{G \mu m_{p}} \times \left(\frac{r}{r_{c}}\right)^3 \times \left(1 + \left[ \frac{r}{r_{c}} \right]^{2} \right)^{-1}    M_{\odot}
\end{equation}

Assuming an isothermal profile for galaxy clusters, r$_{\Delta}$ is given by \citet{LimaNeto}:
\begin{equation}
 r_{\Delta} = r_{c} \left( \frac{2.3 \times 10^8 \beta <kT>}{\Delta h_{70}^2 f^2(z, \Omega_{M}, \Omega_{\Lambda}) \mu r_{c}^2} \right) ,
\end{equation}
where $\beta$ is the shape parameter, $r_{c}$ is the core radius given in kpc, $<kT>$ is the mean cluster temperature given in keV, and $f^{2}$(z, $\Omega$$_{M}$, $\Omega$$_{\Lambda}$) is the redshift evolution of the Hubble parameter.
Accordingly, the total X-ray mass of a galaxy cluster can be estimated from the scaling relations for comparison. Simulations and observations were confirmed the consistency of this approach \citep[e.g.,][]{Ascasibar,Vikhlinin,Vikhlinin2009}. The total mass of galaxy clusters is given by \citet[][]{Vikhlinin}:
\begin{equation}
     H(z) \times M_{2500} =1.25\pm{0.05} \times 10^{14} \times \left( \frac{T_{x}}{5keV} \right)^{1.64\pm{0.06}}    M_{\odot} ,
\end{equation}
where $H(z)$ is the Hubble parameter at redshift z. The $H(z)$ can be estimated from the following equation \citep{LimaNeto}:

\begin{equation}
     H(z) = H_{0} \times f(z, \Omega_{M}, \Omega_{\Lambda}).
\end{equation}

The calculated total masses from the dynamical model and $M$-$<kT>$ relation are found to be highly consistent for our sample of cluster, and the estimated masses are presented in Table \ref{properties}.

\begin{table}
\caption{\label{properties} The physical properties of A3733 system. The $\star$ mark represents galaxy cluster mass estimations by adopting dynamical scaling relations, whereas the $\diamond$ mark represents galaxy cluster mass results obtained from the beta model.}
  \resizebox{0.48\textwidth}{!}{\begin{minipage}{0.54\textwidth}
\begin{tabular}{cccc} 
\hline
\hline

Parameter			& A3733	& 	A3733E				&	A3733W			\\
\hline
z	 			&	0.0380	& 	0.0368$\pm$0.0002	&	0.0394$\pm$0.0002 	\\
V$_r$ (km s$^{-1}$)	&	11405	& 	11044$\pm$45			& 	11825$\pm$45		\\
M$_{2500}^{\star}$  (10$^{13}$  M$_\odot$)		& 11.19$^{+1.04}_{-0.99}$ 	&	4.85$^{+0.62}_{-0.59}$ 	& 	6.34$^{+0.85}_{-0.79}$ 	\\
M$_{2500}^{\diamond}$  (10$^{13}$  M$_\odot$)	& 	14.35$^{+3.17}_{-2.72}$ &	6.24$^{+1.67}_{-1.46}$  &	8.11$^{+2.70}_{-2.29}$ 		 	\\

\hline
\end{tabular}
\end{minipage}}
\end{table}


\begin{figure}
 \begin{center}
 \includegraphics*[width=8.4cm]{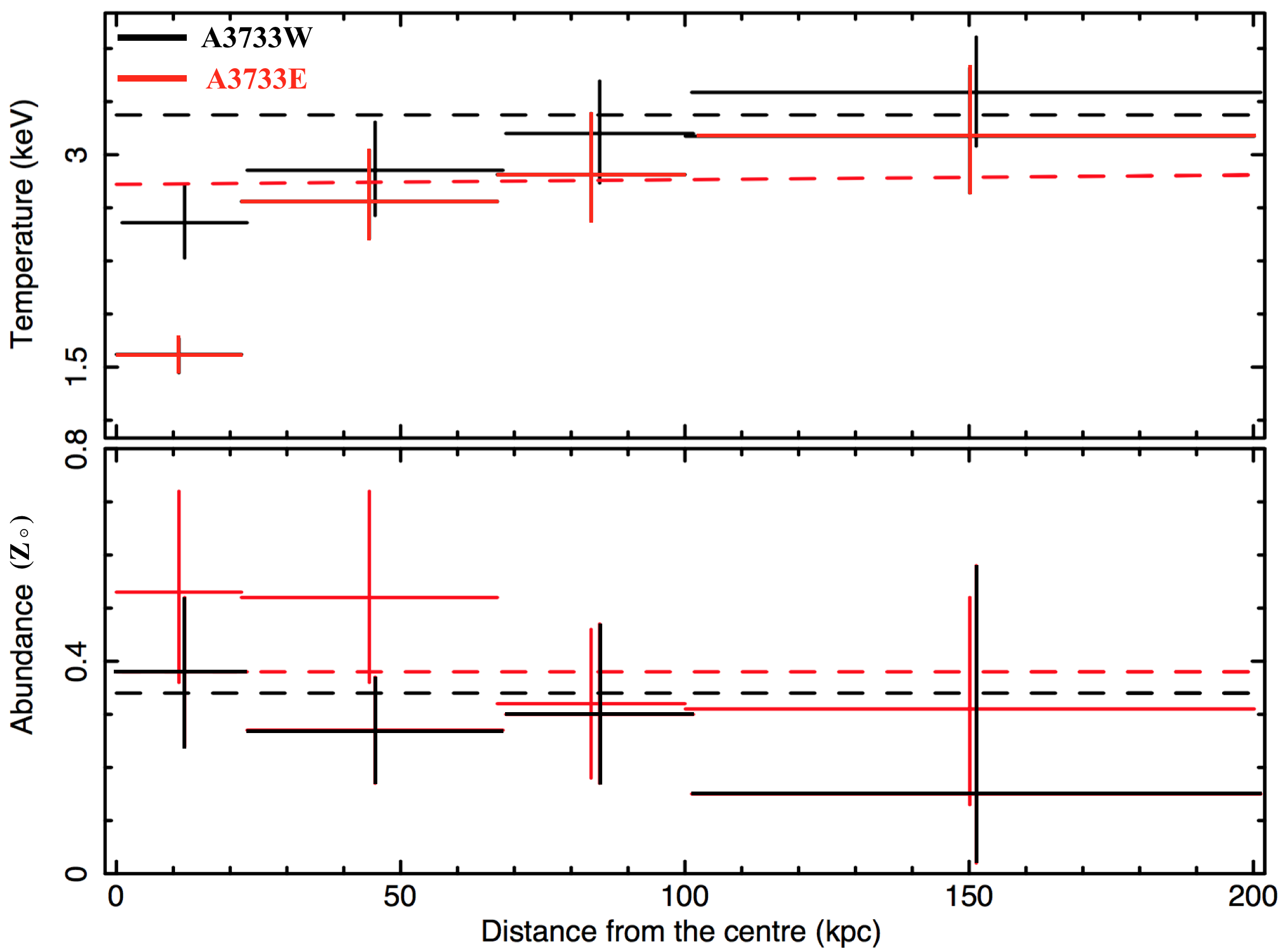}
\caption{\label{spect}The spectral best-fit parameters of the radial profile of A3733, which were obtained by adopting four annular regions. }
\end{center}
\end{figure}

\section{Discussion}
\subsection{Sub-structures}

The X-ray surface brightness map reveals the existence of two separated dense gas clumps in the A3733's cluster vicinity (Fig. \ref{back-sub-im}). The eastern clump is concentrated on BCG1, whereas the western clump is concentrated on BCG2. Interestingly, optical galaxy density map demonstrates that the member galaxies are clustered around BCG2, and there is no nearby companion of BCG1 within 110 kpc radius (Fig. \ref{optical-density} and Fig. \ref{optic}. The galaxy velocity histogram of A3733 is found to be roughly Gaussian (Fig. \ref{optical-density}). X-ray and optical centroids of both BCGs coincide with each other with small positional offsets ($<$ 1kpc). The lack of positional offset between X-ray and optical centroids indicates a pre-merger scenario for A3733, and the cores of both sub-structures have not yet experienced a close core passage.

We additionally studied the surface brightness profiles of each sub-structures using the beta profile (Fig. \ref{sur-bri-im}). Previous X-ray investigations of galaxy groups and clusters have shown that galaxy clusters tend to have $\beta$ > 0.5 \citep[e.g.,][]{Mohr1999}, whereas galaxy groups $\beta$ $<$ 0.5 \citep[e.g.,][]{Mulchaey}. The total X-ray mass of both structures are found to be M$_{2500}$ = 6.24$^{+1.67}_{-1.46}$ 10$^{13}$ M$_\odot$ for A3733E and M$_{2500}$ = 8.11$^{+2.70}_{-2.29}$ 10$^{13}$ M$_\odot$ for A3733W (see Table \ref{properties}. By considering low beta ($\beta$ $<$ 0.5) and core radius ($r_{c}$ $<$ 9 kpc) and the low total X-ray mass, we identify both sub-structures as groups (Table \ref{surbri}). We additionally identify A3733 as a small galaxy cluster due to its low X-ray total mass M$_{2500}$ = 1.44$^{+0.32}_{-0.27}$ 10$^{14}$ M$_\odot$.

The mean ICM temperatures are estimated 2.79$^{+0.21}_{-0.22}$ keV for A3733E and 3.28$^{+0.27}_{-0.25}$ keV for A3733W, whereas the average abundance values are found to be 0.38$^{+0.14}_{-0.12}$ and 0.34$^{+0.13}_{-0.11}$ for A3733E and A3733W, respectively. The ICM temperature of connecting region is found be slightly increased ($kT$ = 4.14 keV). From the X-ray surface brightness distribution, a slight increase can be seen; however, there is still an absence of strong X-ray gas in between sub-structures (see Fig. \ref{sur-bri-im} and Table \ref{spectable}). Therefore, we did not find a direct evidence of adiabatic compression of the gas.

The results of numerical simulations show a high temperature bar, which is almost perpendicular to the collision axis during the pre-assembling stages in the case of low impact parameters. However, the origin of hot region is not certain for pre-merger galaxy clusters. The hydrodynamic processes of heating a plasma are based on the viscous dissipation of energy in a form of the shock or adiabatic compression \citep{book4,S93,T99,RS}.  After the collision, the shock waves are mainly seen in the direction of the original collision axis and extending outwards \citep{book}. The high speed shock waves are only observed with strong surface brightness discontinuities in the post-merger galaxy clusters \citep[e.g,][]{Markevitch2002,Botteon,Dasadia}; however, such strong discontinuities have not seen in the pre-merger galaxy clusters \citep[e.g.,][]{Caglar2}. For the A3733 system, we speculate that the hot region in between sub-structures can be due to shock-heating, even though we did not see any strong discontinuity in the surface brightness profile. But, we note that further investigations are required to understand the origin of hot region in between sub-structures.

To understand the disturbance level of gas, we separately estimate the X-ray morphological parameters for both sub-structures. Due to the high centroid shifts ($w$ $>$ 0.035), we identify both sub-structures as moderately disturbed systems. These results indicate that the merging process has already started. Additionally, the resulting X-ray concentration parameters are found be to c $\sim$ 0.3 for each sub-structure. \citet{Santos} reports that cool core hosting galaxy clusters tend to have $c$ $>$ 0.15. Due to high X-ray morphological parameters of A3733E ($c$ = 0.33) and A3733W ($c$ = 0.29), we identify both sub-structures as cool cores. Additionally, the projected X-ray temperature profiles of A3733E and A3733W demonstrate the existence of cool centres relative to their surroundings (see Fig. \ref{spect}). These results also imply that the centres of both sub-structures have not yet experienced any crucial events that can destroy or dislocate their cool centres. Therefore, we identify the galaxy cluster A3733 as a pre-merger candidate. We present the X-ray morphological parameters results in Table \ref{morpho}.

\subsection{The Dynamical Model of  The A3733 System}

The Newtonian gravitational binding criterion can be applied to understand binding state of A3733. The criterion gives binding probabilities of two-body systems and estimates approaching speed, collision time and relative distance between two-body systems. This model successfully explained binding state of different two-body systems in the literature \citep[e.g.,][]{Beers,cortese,hwang,peng,andreda,nash,Bulbul2016,Caglar1}. The solution of the Newtonian binding criterion describes the binding state of a system in four solutions: two bound incomings (BI), a bound outgoing (BO) and an unbound outgoing (UO).     

The Newtonian criterion for gravitational binding of two-body systems can be solved from the following equation:
\begin{equation}
    V_r^2 \ R_p  \leq \ 2GM \ sin^2 \alpha \ cos\alpha ,
    \label{Newtonian}    
\end{equation}
where $V_{r}$ is the radial velocity difference, $R_{p}$ is the projected separation, $G$ is the gravitational constant, and $\alpha$ is the projection angle. The parametric equation solves the projection angle $\alpha$ for each radial velocity difference $V_{r}$. The radial velocity and the projected separation are related to the projection angle of system:
\begin{equation}
V_r = V sin\alpha, R_p = R cos\alpha ,
\end{equation}
where V and R are the three dimensional velocity difference and separation of the two-body system in the field of sky, respectively.
The parametric motion equation of bound system can be estimated from the following equations:
\begin{equation}
    t = \bigg(\frac{R_m^3}{8GM}\bigg)^{1/2}   (\chi-sin \ \chi),
\end{equation}
\begin{equation}
    R = \frac{R_m}{2}(1-cos \ \chi),
\end{equation}
\begin{equation}
    V = \bigg(\frac{2GM}{R_m}\bigg)^{1/2}    \frac{sin \ \chi}{(1-cos \ \chi)},
\end{equation}
where $R$ is the separation at time $t$, $R_m$ is the separation at the maximum expansion, 
$M$ is the total mass of the system, and $\chi$ is the development angle.

For gravitational unbound systems:
\begin{equation}
    t = \frac{GM}{V_\infty^3}   (sinh \ \chi - \chi),
\end{equation}
\begin{equation}
    R = \frac{GM}{V_\infty^2}(cosh \ \chi -1),
\end{equation}
\begin{equation}
    V =  V_\infty   \frac{sinh \ \chi}{(cosh \ \chi -1)},
\end{equation} 
where $V_\infty$ is the expansion velocity at the asymptotic limit.

The relative probabilities of the solutions can be estimated using the formula:
\begin{equation}
    p_i = \int_{\alpha_{inf,i}}^{\alpha_{sup,i}}  \ cos\alpha \ d\alpha ,   
\end{equation}
where i represents i-th solution of possible binding scenarios. The resulting probabilities were normalised by $P_i = p_i/(\sum_i p_i)$ in our calculations.

\begin{figure}
 \begin{center}
 \includegraphics[width=8.4cm]{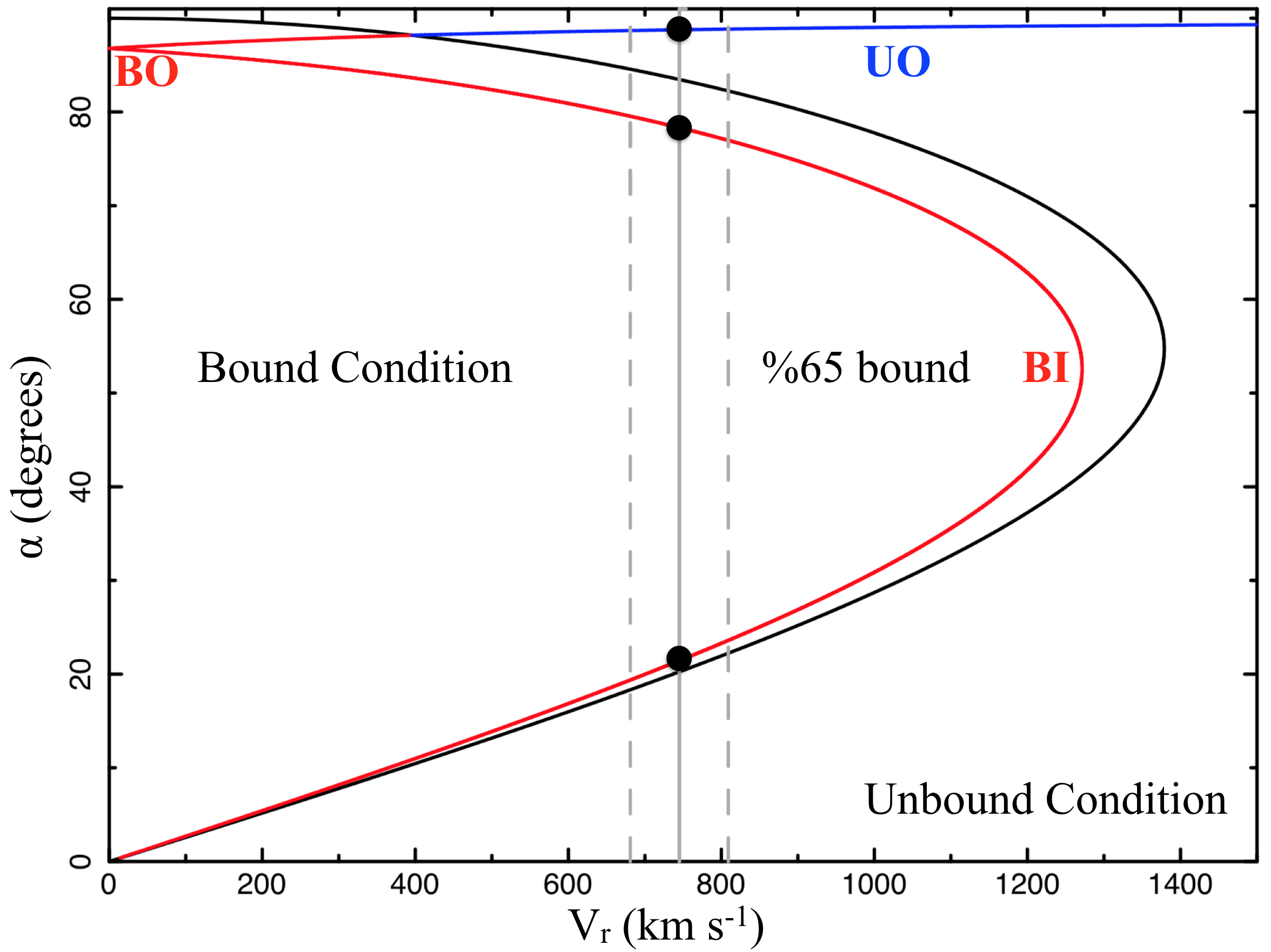}
 \caption{\label{dynamic} The projection angle ($\alpha$) as a function of the relative radial velocity difference ($V_r$) of the sub-structures A3733W and A3733E.
BI, BO and UO stand for Bound-Incoming, Bound-Outgoing, and Unbound-Outgoing solutions. 
The Red and the blue lines represent the bound and the unbound solutions, respectively. 
The black curve represents the limit of bound solutions due to the Newtonian criterion. 
The gray vertical lines demonstrates the relative radial velocity $V_r$ = 781 $\pm{64}$ km s$^{-1}$.}
\end{center} \end{figure}

The following parameters of A3733 were used in the solutions: the radial velocity difference of $V_r$ = 781 $\pm{64}$ km s$^{-1}$, the projected distance $R_p$ = 250 kpc, a total mass of M$_{200}$ = 1.435 $\times$ 10$^{14}$ M$_\odot$, and the age of the universe at cluster redshift $t$ = 12.96 Gyr (4.09$\times$ $10^{17}$ $s$). The resulting solution of the Newtonian gravitational binding criteria of A3733 is presented in Fig. \ref{dynamic}. Solving the equation \ref{Newtonian}, the system is found to be \% 65 bound. In addition, the parametric solution gives three solutions for A3733: two bound in-comings and an unbound outgoing (see Fig. \ref{dynamic}). The first solution (\% 11.7 probability) indicates that A3733E and A3733W have 3D separation R = 1.16 Mpc and approach each other with colliding velocity V = 763 km s$^{-1}$. The high probable (P = \% 88.2) scenario suggests that the cores of sub-structures have a 3D separation of 0.27 Mpc and will collide in 0.14 Gyr with the relative in-falling velocity of 1936 km s$^{-1}$. The unbound scenario (\% 0.01 probability) results in 3D separation of 12.0 Mpc. The parameters of parametric solutions are presented in Table \ref{bound} and \ref{unbound}. We note that, as mentioned in \citet{andreda}, the solution of this method does not involves angular momentum of the system due to absence of angular momentum information of the colliding structures.

\begin{table}
\begin{center}
\caption{\label{solution} The estimated parameters for bound incoming solutions of the two-body dynamical model. }
\begin{tabular}{@{}cccccc@{}} 
\hline
\hline
 $\chi$	& 	$\alpha$	&	$R$ 		&	$R_m$	& 	$V$			&	$P$	\\	
 (rad)	&	(degrees)	&	(Mpc)	&	(Mpc)	&	(km s$^{-1}$)	&	(\%) 	\\
\hline
4.92 		&	77.58	& 	1.16		&	2.92	&	763 		&	11.7 \\
5.65		&	22.63	& 	0.27		&	2.81	&	1936		&	88.2 \\
\hline
\hline
\end{tabular}
\label{bound}
\end{center}
\end{table}

\begin{table}
\begin{center}
\caption{\label{unbsol} The estimated parameters for unbound outgoing solution of the two-body dynamical model.}
\begin{tabular}{@{}cccccc@{}} 
\hline
\hline
 $\chi$	& 	$\alpha$	&	$R$ 		& 	$V$			&	$V_\infty$		&	$P$	\\	
 (rad)	&	(degrees)	&	(Mpc)	&	(km s$^{-1}$)	&	(km s$^{-1}$)	&	(\%) 	\\
\hline
3.08 		&	88.81	& 	12.0	&	745		&	680		&	0.01 \\
\hline
\hline
\end{tabular}
\label{unbound}
\end{center}
\end{table}

\subsection{Brightest Cluster Galaxies}

A3733 hosts two binary galaxy groups, which are concentrated on two different BCGs (NGC 6998 for A3733W and NGC 6999 for A3733E). The velocity histogram of galaxies can be fitted with a single Gaussian model; however, the region around NGC 6998 appears to be crowded with member galaxies, whereas NGC 6999 has no companion within 110 kpc radius (Fig. \ref{optic} and Fig. \ref{optical-density}). The X-ray centroids of both BCGs coincide with the optical centroids with a displacement of less than 1 kpc. The X-ray and optical association are estimated using the following equation, which is explained in detail by \citet{Pineau}

\begin{equation}
LR (r) = \frac{1}{2\lambda} e^{-0.5r^{2}},
\end{equation}
where LR is the likelihood ratio of association, $\lambda$ is the pilot function, r is the distance between X-ray and optical centroids. The resulting LR values are found to be \% 96, \% 88 for NGC 6998 and NGC 6999, respectively.

\subsubsection{NGC 6998}

NGC 6998 is the second brightest galaxy in the vicinity of A3733 \citep{PoLa}. The diffuse ($\sim$ 400 kpc) radio emission, which comes from the central black hole is associated with NGC 6998 \citep{Morganti}. In Figure \ref{Radio}, the radio lobes and a strong jet can also be seen from the $GMRT$ 150 MHz and $VLA$ 1400 MHz image. The X-ray spectrum of NGC 6998 was fitted with thermal and power law component by adopting a redshift dependent fixed nH density of this source ([apec+powerlaw]*ztbabs). Based on the fit, we obtained the X-ray luminosity $\log{L_{x}}$ = 41.75 erg s$^{-1}$ and the hardness ratio HR = 0.55 $\pm{0.04}$. We note that the hardness ratio is estimated using the following formula HR = H-S/H+S, where H is count rate in 2.0-10.0 keV and S is count rate in 0.5-2.0 keV \citep{Caglar0}. Therefore, we identify this source as the low luminosity active galactic nuclei. 

\subsubsection{NGC 6999}

NGC 6999 is the brightest galaxy of A3733 \citep{PoLa} and it has no nearby companion within 110 kpc radius. The X-ray spectra of this source is well-modelled with a single thermal model with a fixed $nH$ density value (apec*tbabs). The resulting X-ray luminosity is obtained as $\log{L_{x}}$ = 40.80 erg s$^{-1}$ for this source. 

\section{Summary}
In this study, X-ray and optical data were used to perform structural analysis of A3733. The obtained features indicate that A3733 is a bimodal system, which is in an early stage of merger. The main findings of the study are summarized as follows: 
\begin{itemize}

\item The optical density map reveals the existence of dense region around the BCG2. Moreover, the velocity histogram of member galaxies results in two peak points, which gives an impression of clustering around both BCGs (see Fig. \ref{optical-density}). 

\item A3733E is concentrated at the BCG1, and A3733W is concentrated at the BCG2. While X-ray centroid of the BCG1 is dislocated from optical centroid by $\sim$ 1 kpc, X-ray centre of the BGC2 matches with its optical centre (see Fig. \ref{optic}).  

\item The average temperature values of A3733E and A3733W are found to be 2.79 keV and 3.28 keV, respectively (see Table \ref{spectable}).  

\item The both sub-structures are found to host cool cores (see Fig. \ref{spect}, Table \ref{morpho}). 

\item The mass calculations result in M$_{2500}$ = 6.24 $\times$10$^{13}$ M$_{\odot}$ for A3733E and M$_{2500}$ = 8.11 $\times $10$^{13}$ M$_{\odot}$ for A3733W (see Table \ref{properties}).

\item The surface brightness profile of A3733 reveals an absence of strong X-ray emitting gas between sub-structures (see Fig. \ref{sur-bri-im}).

\item The high probable (\% 88.2) dynamical binding model suggest that the cores of sub-structures have a 3D separation of 0.27 Mpc and will collide in 0.14 Gyr with the relative in-falling velocity of 1936 km s$^{-1}$ (see Table \ref{bound}).

\end{itemize}

In this work, we find some pieces of evidence that A3733 is a pre-merger system due to the following signatures; a) X-ray and optical centres of BGCs substantially coincide with each other, b) the gas within sub-structures' cores are not disturbed, c) there is an absence of a strong X-ray emitting gas between two sub-structures, d) both sub-structures are found to host cool cores, e) the ICM temperature between sub-structures is slightly increased (kT = 4.14 keV). In conclusion, the system is clearly at the early stage of the merger. For future research, this study encourage a longer exposure X-ray observation and comparisons with numerical simulations.

\section*{Acknowledgement}

Authors would like to thank anonymous referee for his valuable comments and suggestions. Authors are grateful to G\"{o}zde \"{O}zzeybek for her feedbacks and comments. TC also would like to thank Reinout van Weeren, Aurora Simionescu, Gabriella Di Gennaro, Dilovan Serindag and Henk Hoekstra for very useful discussions.



\appendix

\section{Radio Images}

\begin{figure}
 \begin{center}
\includegraphics*[width=6.6cm]{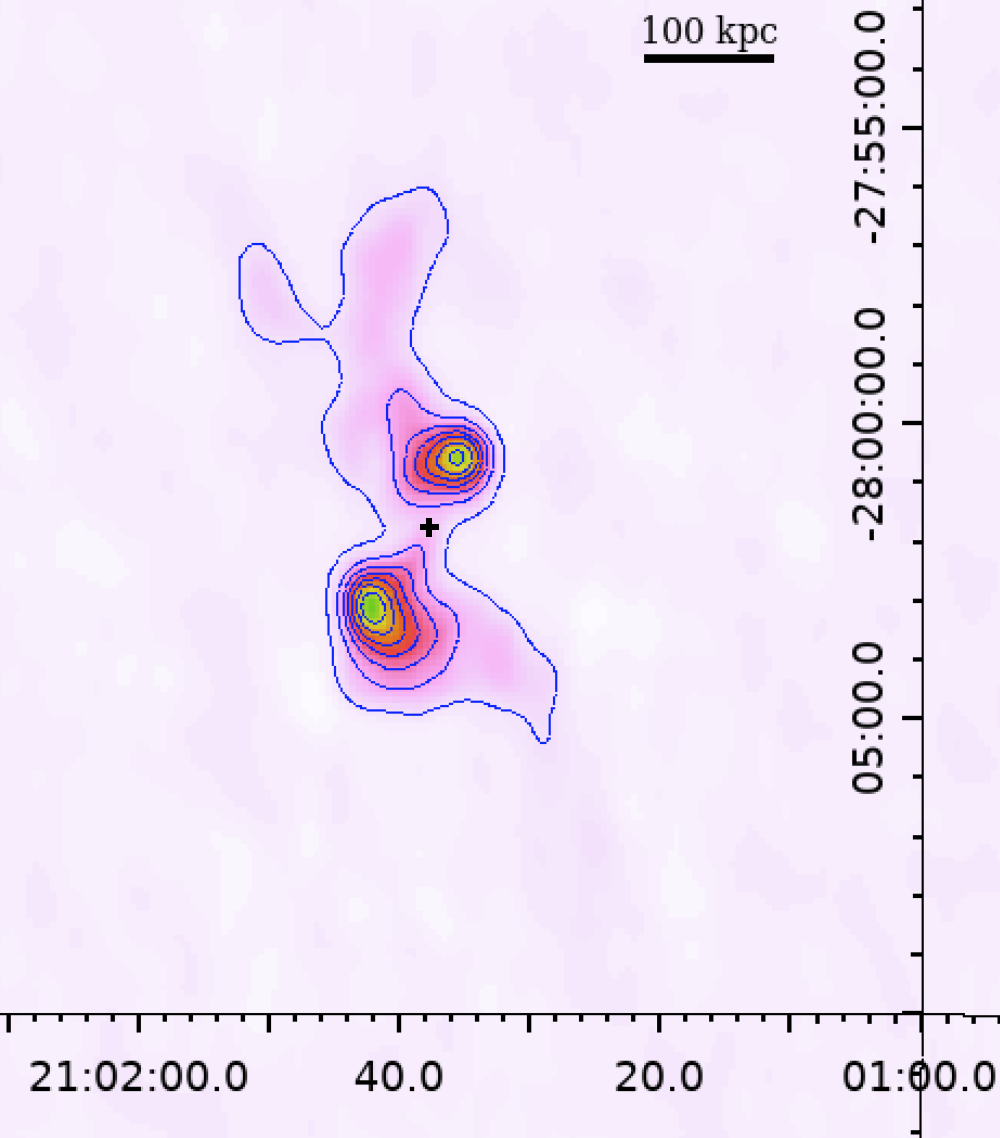}
\includegraphics*[width=6.6cm]{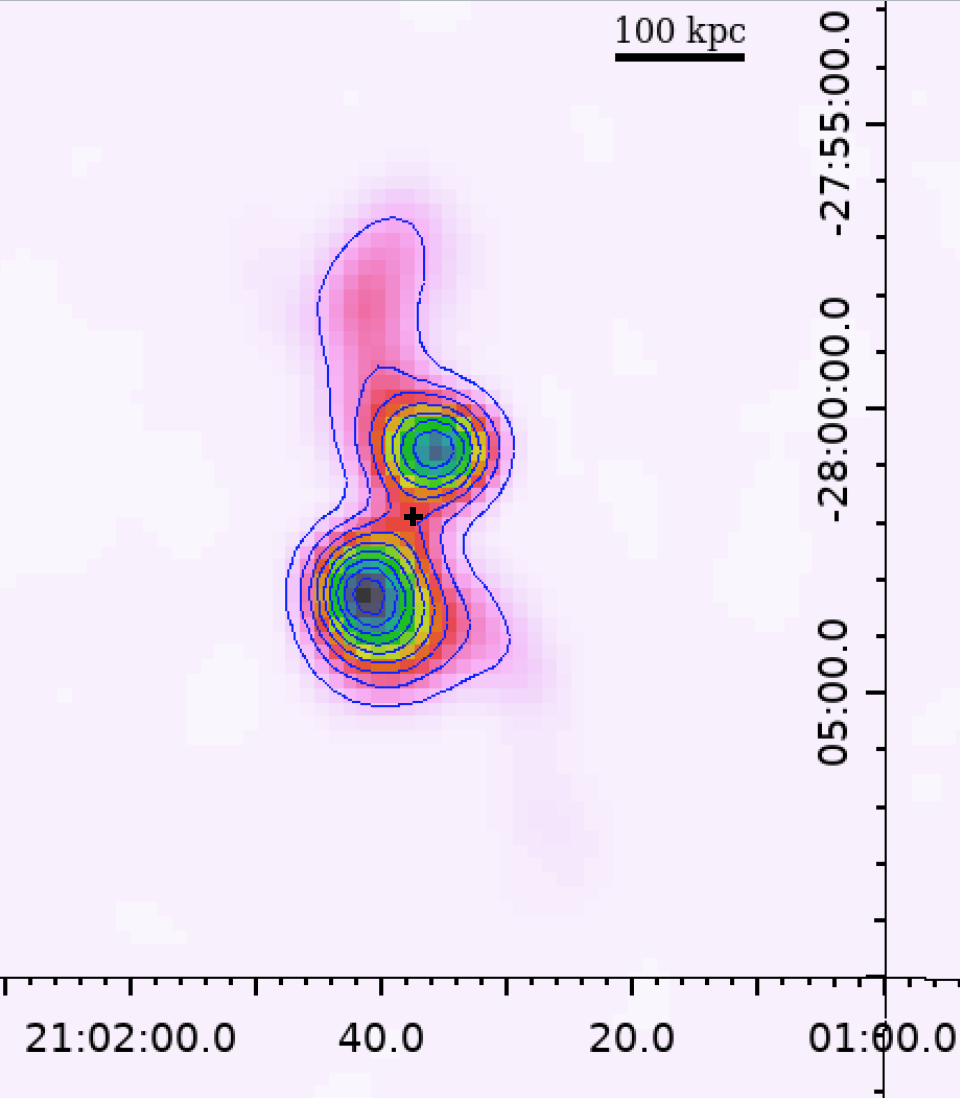}
\includegraphics*[width=6.6cm]{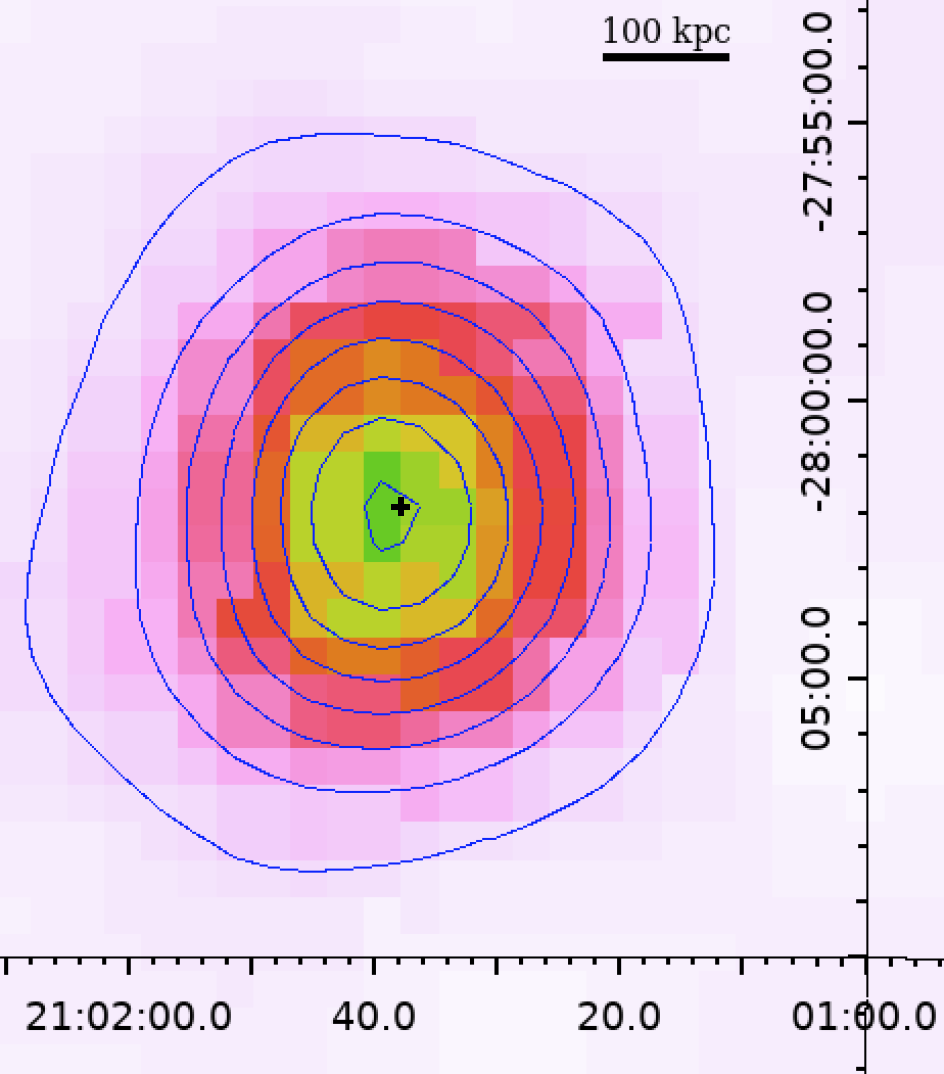}
 \caption{\label{Radio} The radio data $\textbf{Top:}$ GMRT 150 MHz all-sky radio survey; $\textbf{Middle:}$ The NRAO VLA 1400 MHz sky survey; $\textbf{Bottom:}$ The $Green$ $Bank$ 4850 MHz sky survey. The black plus signs demonstrate the position of BGC2.}
\end{center} \end{figure}

We present the radio images, which are obtained from the $GMRT$ 150 MHz all-sky radio survey, The $NRAO$ $VLA$ 1400 MHz sky survey and The $Green$ $Bank$ 4850 MHz sky survey. The absence of diffuse synchrotron emission in the form of radio halo or relics, which were only observed in post-merger galaxy clusters \citep[e.g.,][]{DiGennaro}, also supports pre-merger scenario indirectly for A3733 (see Fig. \ref{Radio}).

\bsp	
\label{lastpage}
\end{document}